\documentclass[12pt]{spieman}  % 12pt font required by SPIE;

\usepackage{amsmath,amsfonts,amssymb}
\usepackage[numbers]{natbib}
\usepackage{graphicx}
\usepackage{setspace}
\usepackage{tocloft}
\usepackage[dvipsnames]{xcolor}
\usepackage{graphicx}
\usepackage{longtable}
\usepackage{multirow}
\usepackage{lineno}

\newcommand{\Lya}{Ly$\alpha$}
\newcommand{\Lyb}{Ly$\beta$}
\newcommand{\Lyg}{Ly$\gamma$}
\newcommand{\Lyd}{Ly$\delta$}

\newcommand{\HI}{H~{\sc i}}

\newcommand{\OI}{O~{\sc i}}

\newcommand{\OVI}{O~{\sc vi}}
\newcommand{\SII}{S~{\sc ii}}

\newcommand{\FeII}{Fe~{\sc ii}}
\newcommand{\SiII}{Si~{\sc ii}}
\newcommand{\SiIII}{Si~{\sc iii}}
\newcommand{\SiIV}{Si~{\sc iv}}
\newcommand{\CaII}{Ca~{\sc ii}}
\newcommand{\CII}{C~{\sc ii}}
\newcommand{\CIII}{C~{\sc iii}}
\newcommand{\CIV}{C~{\sc iv}}

\newcommand{\NV}{N~{\sc v}}

\newcommand{\NeVIII}{Ne~{\sc viii}}
\newcommand{\MgII}{Mg~{\sc ii}}
\newcommand{\CrII}{Cr~{\sc ii}}
\newcommand{\ZnII}{Zn~{\sc ii}}

\newcommand{\vlsr}{v_{\rm LSR}}
\newcommand{\kms}{km~s$^{-1}$}

\newcommand{\msunyr}{${\rm M_\odot~yr^{-1}}$}

\title{Characterizing gas flows through observations of the disk-CGM interface with the HWO}

\author[a]{Sanchayeeta Borthakur}
\author[b]{Joseph N. Burchett}
\author[c]{Frances Cashman}
\author[d]{Andrew J. Fox}
\author[e]{Yong Zheng}
\author[f]{David M. French}
\author[g]{Rongmon Bordoloi}
\author[a]{Brad Koplitz}

\affil[a]{School of Earth \& Space Exploration, 
Arizona State University, 781 Terrace Mall, Tempe, AZ 85287, USA}
\affil[b]{Department of Astronomy, New Mexico State University, Las Cruces, NM 88003, USA}
\affil[c]{Department of Physics, Presbyterian College, 503 South Broad Street
Clinton, South Carolina 29325}
\affil[d]{AURA for ESA, Space Telescope Science Institute, 
3700 San Martin Drive, Baltimore, MD 21218, USA}
\affil[e]{Department of Physics, Applied Physics and Astronomy, Rensselaer Polytechnic Institute, Troy, NY 12180, USA}
\affil[f]{Space Telescope Science Institute, 3700 San Martin Drive, Baltimore, MD 21218, USA}
\affil[g]{Department of Physics, North Carolina State University, Raleigh, NC 27695, USA}

\cftpagenumbersoff{figure}
\cftpagenumbersoff{table} 

\begin{document} 
\maketitle

\begin{abstract}
How gas gets into, through, and out of galaxies is critical to understanding galactic ecosystems. The disk-CGM interface region is uniquely suited for studying processes that drive gas flows.  
Matter and energy that enter and leave a galaxy pass through this region; however, the precise pathways are yet to be explored. 
In this paper, we discuss future observations that will facilitate the discovery of the gas flow pathways in galaxies and the telescope parameters necessary for making those observations. 
We advocate for high spectral resolution ultraviolet spectroscopic capabilities on the Habitable Worlds Observatory (HWO) that will {\color{black} enable observations at a wavelength range of 940--3500 \AA\ (minimum range 970--3000 \AA) and at a resolution of 100,000 (minimum of 50,000). 
We advocate for a multi-object spectrograph with thousands of sub-arcsec slitlets and a field of view $\rm 6^{\prime} \times 6^{\prime}$. }
We also recommend that the spectrograph be sensitive enough to achieve a signal-to-noise ratio of 10 or higher within a few hours for a continuum source of 21 AB magnitude {\color{black} and estimate an optimal aperture size of 8 meters.}
These capabilities would enable the characterization of gas in the disk-halo interface, leading to breakthroughs in our understanding of the gas flows and galactic ecosystems.

\end{abstract}

% Include a list of up to six keywords after the abstract

% Include email contact information for the corresponding author
{\noindent \footnotesize\textbf{*}Sanchayeeta Borthakur,  \linkable{sanchayeeta.borthakur@asu.edu} }

\begin{spacing}{2}   % use double spacing for rest of manuscript

\section{Introduction \label{sec:intro} } % \label{} allows reference to this section

Over the last few decades, a general picture has emerged from simulations of how cool gas flows into and out of star-forming galaxies. In this picture, cool-gas inflows occur primarily along or near the disk plane, while outflows occur predominantly along the minor axis, perpendicular to the disk \citep{keres2005, stewart2011}. While this picture is widely accepted and is supported by observational evidence that CGM properties depend on azimuthal angle \citep{bordoloir2011, Kacprzak2012} it has not been tested in detail and does not account for the complexity of the disk-halo interface. Nor does this model account for the locations where {\it hot gas} cools and accretes onto galaxies \citep{Fraternalli2017}, which may be more isotropic. 

Studies of the Milky Way (MW) have shown that the disk-halo interface is far more complicated and structured than this simple model predicts because cool-gas inflows and outflows are found in different locations across the sky rather than only along the major and minor axes \citep{Richter2017, fox19}. Therefore, our picture of the inflow-outflow pathways around star-forming galaxies needs much more data from galaxies across a range of redshift and from multiple locations within single galaxies before we can claim to understand how matter flows into and out of galaxies.

Gaining a census of gas flows, their origins, and their fate is one of the critical aspects highlighted in the 2020 Astronomy Decadal Survey under the priority area “Unveiling the Hidden Drivers of Galaxy Growth” and one of the key scientific recommendations for a large-aperture ultraviolet(UV)/optical space telescope (Astro2020, \S2.3.3 ). 

The science case presented here is specifically designed to highlight observations that will provide answers to the question, “How does matter flow into and out of galaxies and fuel their growth?” To do so, we need to examine the interface region between the galaxy disk and its halo, also referred to as the circumgalactic medium (CGM).

\subsection{The Key Science Questions for the Habitable Worlds Observatory}
The scientific objective of this science case for the Habitable Worlds Observatory (HWO) is to characterize the distribution of gas and metals in the disk-CGM interface to discover the pathways through which gas, metals, and dust flow into and out of galaxies (Astro2020, D-Q2). 
We propose the following scientific tasks. In {\color{black}brackets}, we highlight the sub-questions discussed in Astro2020 that these specific tasks will answer.

\begin{itemize}
    \item[1.] Characterize the distribution and kinematics of gas clouds beyond the {\color{black}traditionally regarded} gas disk {\color{black} and into} the disk-CGM interface ($\sim$10\% of the virial radius) [D-Q2a] to measure the gas mass available to support future star formation. 
    \item[2.] Quantify the total gas flow rates - towards and away from the disk - to understand how galaxies acquire the gas necessary to fuel star formation [D-Q2b] and lose gas from the disk through feedback processes.
    \item[3.] Measure metal content and abundance of elements to trace the cycling of metals outside of the stellar disk [D-Q2b].
    \item[4.]Determine energetics and kinematics of the gas in the disk-CGM interface in various phases over a range of temperatures that enable assessing the coupling of energetic feedback processes to the gas in the disk-CGM interface [D-Q2c].

\end{itemize}

The next-generation UV flagship mission should provide spectroscopic capabilities with sensitivity and resolution to detect and measure gas properties (physical, chemical, ionization, and kinematic) in the disk-CGM interface [D-Q2d]. 

Here, we begin by discussing the current understanding of gas flow processes from various observations at the disk-CGM interface and the science objective for future observations in \S\ref{sec:gas_flows}.  We describe the measurements and derived physical parameters needed to make advancements in \S\ref{sec:physical_parameters}. Guided by the science motivators, we make specific recommendations for instrument design for HWO in \S\ref{sec:HWO_performance}. We summarize our recommendation in \S\ref{sec:conclusion}.

\section{Overview and Science Objectives\label{sec:gas_flows}}

The disk-CGM interface hosts signatures of various processes that bring and remove gas from the disk. Here, we highlight five avenues of scientific investigation that will enable the direct detection of gas flows into, through, and out of the disks of galaxies.

\subsection{ Extended disks \label{extended_disks}}

Most star-forming galaxies have gas disks that are at least three times larger than their stellar disk \citep{swaters02}. Traditionally, the radius of the \HI\ disk is defined as a region where the gas column density is $\rm 1~M_{\odot}~pc^{-2} \equiv 2 \times 10^{20}~cm^{-2}$, however, the gas disks of galaxies go much further, albeit at lower column densities. This region is commonly known as the extended disk \citep{kalberla09}. The column density in the extended disks tends to drop exponentially as a function of galactocentric radii and reaches limits that are not measurable in \HI\ 21cm emission ($\sim \rm few \times 10^{18}~cm^{-2}$). QSO-absorption line spectroscopy, especially tracing UV-absorption lines, is ideal for detecting faint gas in the extended disk that is otherwise not detectable.

The presence and properties of the extended disk are critical for testing theories on the pathway of gas flows into galaxies. One proposed pathway for gas flows from the intergalactic medium (IGM) is through the CGM that eventually joins the extended disk and then gradually spirals inward \citep{stewart2011}. This is supported by the fact that the IGM and the CGM surrounding the gas-rich galaxies are rich in \Lya\ absorbers \citep{borthakur2015, borthakur22}. 
The other popular theory of gas accretion is the fountain flow, where previous generations of outflows act as seeds for the CGM to cool onto them and rain vertically along the minor axis of the galaxy \citep{Fraternalli2017}. Recent theoretical studies have found both to be at play simultaneously, although specific predictions vary between simulations. This necessitates quantifying the properties of the extended disk, such as its size, gas column density profile, metallicity, dust content, kinematics, and angular momentum \citep{Stewart17, Renzini20}, which in turn will constrain gas accretion models. 

Extended co-rotating disks at large distances from the galaxy (50-100~kpc) have been detected in Mg~II and other species \citep{Kacprzak10, Ho17, martin19, Diamond-Stanic16, french2020, dupuis21, Nateghi24a, Nateghi24b, olvera24} especially in galaxies at z$>$0.3 where the \MgII\  doublet is accessible from the ground. However, these disks are not ubiquitous, and non-detection of co-rotating disks has also been reported \citep{Ho17, martin19,Nateghi24a}. What allows galaxies to have large co-rotating disks is still debatable. Galaxies at lower redshift, where galaxy properties can be studied in great detail \citep{borthakur2024}, have yet to be thoroughly investigated for extended disks. 

Studies of the extended disk in the nearby Universe will need access to high-resolution far UV-absorption spectroscopy. The full coverage of spectral window from far-UV ($\sim$ 970 \AA) to near-UV (3000 \AA), mainly covering the higher order Lyman series transitions (\Lyb\ and \Lyg), metal-tracers of neutral gas such as \OI $\lambda\lambda \rm 1025, 1039, 1302$, low-depletion metal tracers such as \SII $\lambda\lambda \rm 1250, 1253, 1259$, \CrII $\lambda\lambda 2056, 2062, 2066$, and dust tracers \ZnII $\lambda \lambda \rm 2026, 2062$, will enable us to - (1) characterize the baryonic mass hidden in the extended disk and co-rotating CGM; (2) measure the kinematics and cloud distribution in the gas that sustains the gas disks of galaxies; (3) conduct a census of metals in the extended disk and the inner CGM; and (4) identify signatures of gas inflows and outflows feeding and stripping the disks.

\subsection{High Velocity Clouds (HVCs) \label{HVCs}}

In addition to a rotating disk, galaxies contain a significant fraction of their neutral and ionized gas in clouds above and below the disk \citep{kalberla09}. These clouds are not rotationally supported and show significant velocity offsets from the rotating disk.

In the MW, these clouds are called high-velocity clouds (HVCs).
They have velocities $|\vlsr| \gtrsim90-100$ \kms\ that cannot be explained by corotation with the disk \citep{wakker91}.
Overall the bulk of the HVCs are infalling towards the disk with an estimated inflow rate of 0.5--1.4 \msunyr\ \citep{lehner11, putman12, Richter2017, fox19}. The gas flow rate brought by the HVCs is sufficient to sustain the observed star formation rate of the MW provided that the HVCs survive their infall onto the disk.
In the UV, which provides access to much lower column densities, the HVC covering fraction is $\sim100\%$ at ${\rm N_{\rm HI} \gtrsim 10^{15}~cm^{-2}}$ \citep{french2021}. The dense cores of the HVCs are surrounded by lower column density media that show significant columns in species traced by UV absorption lines such as \SiII, \SiII, \SiIV, \CIV, \OVI\ \citep[e.g,][]{savage93, fox05, wakker12}, and optical absorption lines such as \CaII\ \citep[e.g.,][]{wakker07}.

\begin{figure}
    \centering
    \includegraphics[width=0.99\linewidth]{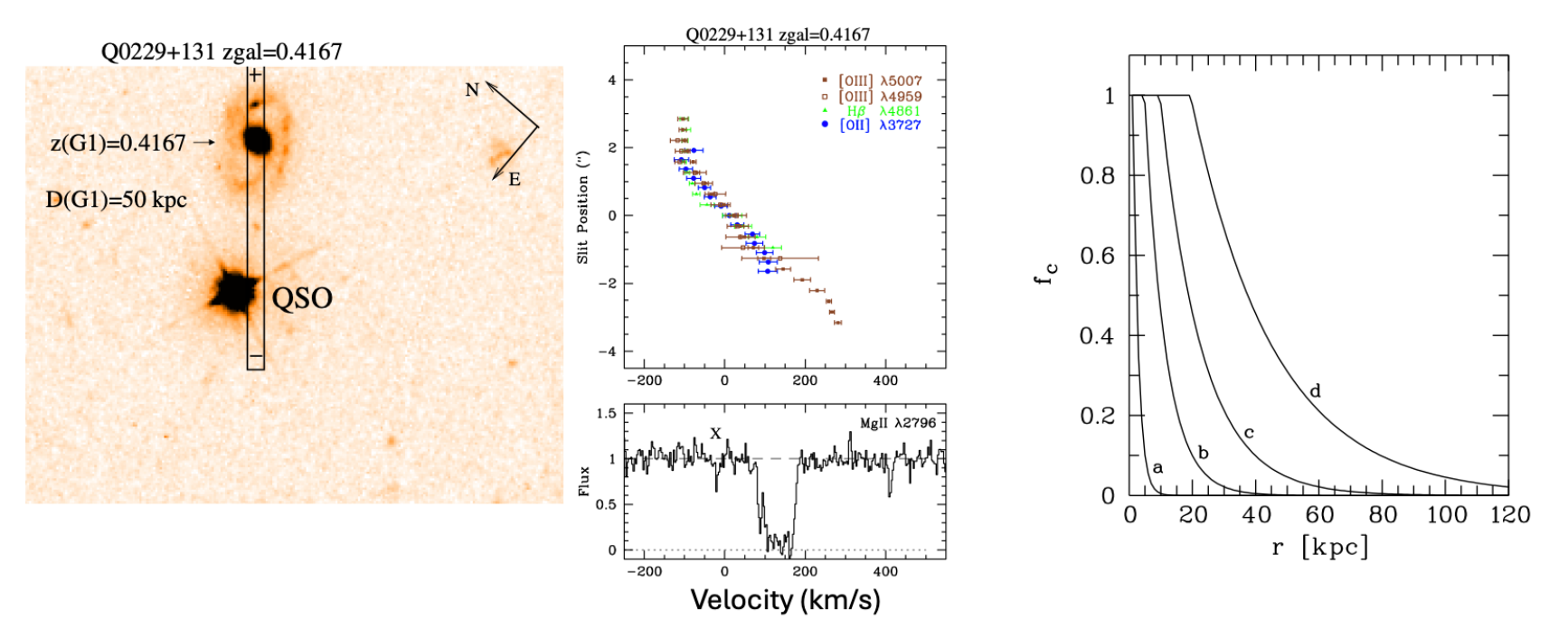}
    \caption{Left: Presence of an extended disk seen as a co-rotating Mg II absorber {\color{black}projected 50 kpc from} a z=0.4167 galaxy (adapted from \citep{Kacprzak10}). The middle panels show the rotation curve (top) and the absorption profile (bottom), suggesting corotation at least twice the radius of the stellar disk. Right: Modeling of the covering fraction of high-velocity clouds for MW-type galaxies. Predictions for galaxies with four different \HI\ masses i.e., log (M(HI)disk)= 7.75(a), 8.95(b), 9.55(c), 10.15(d), are shown. (adapted from \citep{Richter2012}). }
    \label{fig:mol_stack}
\end{figure}

{\color{black} Our understanding of the prevalence of structures like HVCs in other galaxies is limited. Only a handful of extragalactic targets have been explored at the sensitivity levels needed to detect and characterize such non-corotating clouds, often referred to as anomalous velocity clouds (AVCs).  These include M31, M51, M83, M100, M101, NGC~628, NGC~2403, and NGC~6946, among a few others \citep{Thilker04,Westmeier08, Miller05, Miller09, Gim21, vanderHulst88, Fraternali01, Kamphuis_Briggs92, Kamphuis_Sancisi93, Boomsma08}. Although velocity anomalies from rotation are hard to detect for inclined or edge-on galaxies, the extraplanar gas seen in galaxies like NGC~891 and others \citep{Oosterloo07,Heald11} is believed to be analogous to AVCs. Most AVCs were detected by mapping the gas in \HI\ 21cm hyperfine transition and as a result the chemical, physical, and ionization state of the gas cannot be evaluated. }

Absorption spectroscopy with the HWO promises to push the boundaries of understanding of HVCs in the MW and AVCs in external galaxies by providing access to many more UV-bright sightlines and allow detailed characterization of their structure and distribution. Doing so will allow us to answer critical outstanding questions: 1) How prevalent are AVCs in other galaxies? 2) What are the sizes and density structures of these clouds? 3) Are HVCs (AVCs) able to survive their journey to the disk? If so, how? 4) What is the kinematic distribution of the HVCs (AVCs) and how do the populations change as a function of their velocity offset from the disk? 5) What are the variations in metal abundances and molecular content across HVCs and their role in feeding star formation in the disk of their host galaxies?

Answering these questions would require observations that probe HVCs at small scales, enabling us to measure sizes and variations in properties across and within the clouds. Similarly, multiple sightlines within 30 kpc of extragalactic systems are required to quantify whether other galaxies have structures similar to those of MW HVCs in their disk-CGM interface. A key sensitivity requirement for achieving the science is that the HWO can detect continuum sources down to a FUV magnitude of 21 at a signal-to-noise ratio of 10 to 15, making such observations feasible. {\color{black} We discuss more on the sensitivity requirements in the next section.}

\subsection{Molecular Hydrogen in the Disk-CGM interaface \label{molecular_hydrogen}}

A crucial aspect of the baryon budget in the disk-halo interface is accounting for the molecular gas, particularly molecular hydrogen (H$_2$), which can hold a substantial amount of mass compared to other phases.

Past observations of molecular hydrogen with the Far-ultraviolet Spectroscopic Explorer (FUSE) allowed astronomers to convincingly prove the presence of molecular gas in the disk-CGM interface in the MW and a few other galaxies {\color{black}including those undergoing starbursts} \citep{vandeputte2023, shull2021, gillmon2006, bluhm2003,tumlinson2002,bluhm2001, hoopes2004}.
However, due to the current lack of access to the far-UV wavelengths below 1100 \AA\ at medium ($R \sim 1,000$) to high resolution ($R>20,000$), no further progress has been made since FUSE.

HWO is uniquely suited to observe molecular hydrogen in the FUV, which will enable us to probe all phases of the molecule irrespective of the temperature. The FUV gives us access to up to 8 electronic bands (from $v$($0-0$) to $v$($8-0$)) of $H_2$ within the wavelength range from 1000 to 1152~{\AA}. 
Each band hosts multiple rotational transitions ($\sim$14 per band from $J=0$ to $J=7$), which provides us the highest density of H$_2$ transitions, with 157 transitions from 1152 to 1000~{\AA} alone. {\color{black} Furthermore, extending the wavelength down to the 940~{\AA} would enable observations of almost 318 lines.}

Examples of UV-based H$_2$ detections come from FUSE (launched: 1999; mission ended: 2007) observations about two decades ago. These data enabled the first detection of high-velocity H$_2$ associated with the nuclear wind of the Milky Way (MW) \citep{Cashman2021}. More than 60 transitions of H$_2$ from the ground state and excited rotational levels ($J=0$ to 5) in two high-velocity H$_2$ clouds were detected in a region approximately one~kpc below the Galactic Center. Access to UV lines enabled measurement of excitation temperatures in the gas and discovery of the cloud structure, featuring a cold molecular core and a warmer envelope heated by processes such as UV pumping \citep{richter2001}.
Although this result provides a unique window to understand the feedback process in the disk-CGM interface, this is only one sightline where such data were available. The only other molecular hydrogen detections in the disk-CGM interface come from high-velocity clouds (HVCs) \citep{richter1999, richter2001, sembach2001, wakker2006, Tchernyshyov2022} due to FUSE's limited sensitivity to observe bright background QSOs.

The diagnostic purpose of the UV lines differs from that of the near-infrared (1-5 microns) and far-infrared (5-30 microns) lines. The UV lines trace high-energy processes and excitation mechanisms that infrared (IR) observations might miss because the rotational and vibrational bands in the IR only trace the very cold phase. 
Combining UV and IR data will provide a more synergistic approach to understanding the complex microphysics and velocity structure within molecular clouds, with UV H$_2$ lines tracing higher-velocity gas that is more affected by energetic processes.

FUV spectroscopic capability on the HWO will open a window to the unseen frequencies in the local Universe. In the absence of any FUV instrumentation in any current NASA missions, it is paramount that the next-generation UV spectroscopic design includes the rest of the frame molecular hydrogen bands ($\leq$ 1152 \AA) at high spectral resolution required to distinguish individual components at 6~\kms\ (R$>$ 50,000). These requirements will allow us to answer vital outstanding questions such as:
1) How common are H$_2$ outflows at the disk-halo interface?
2) How far from the disk can they survive in the CGM?
3) Are H$_2$ outflows connected to the observed neutral \HI\ and low-ionization metal outflows? \citep{mcclure2013,lockman2016,DiTeodoro2020,LockmanJ2020}.

\subsection{Feedback processes in the disk-CGM interface}

Energetic feedback - the ``catch-all" term that describes the physical processes via which stars or AGNs deposit energy, momentum, and mass into the CGM, is vital in regulating galaxy evolution \citep{somerville2015}. 
Despite their importance, the underlying {\color{black} physical conditions of the outflowing gas are poorly constrained.} Feedback from supernovae or AGNs originates at localized regions in a galaxy disk (at parsec scales), but they can sweep up the local ISM and propagate it out to tens to hundreds of kilo-parsec into the CGM \citep{borthakur2013, bordoloir2011, heckman2017}. 
However, owing to the small scales of the driving sources, no current cosmological simulation can implement these models ``ab initio." As a result, all current cosmological models rely on highly uncertain ``sub-grid" prescriptions of the feedback processes \citep{Crain2023}. The only practical approach to constrain these ``sub-grid" feedback models is to perform direct observations.

The disk-CGM interface is a unique zone where direct detection of outflowing material is possible and may be unambiguously connected to the site of energy injection. In addition, due to the low density of the medium, the ionization and kinematic signatures of the energetic feedback processes can be retained for tens of millions of years, thus enabling us to measure these signatures.

\subsubsection{Gas Outflows and Stellar Feedback Processes}

Traditionally, outflows have been extensively studied through ``down-the-barrel" spectroscopy of star-forming/starbursting galaxies, where outflowing material is detected as blueshifted absorption against the background stellar light of the host galaxy \citep[e.g.;][]{Heckman2015, heckman2016, Bordoloi2014_outflow}. But such observations suffer from two significant limitations: first, the positional information of the outflowing material is lost, and second, the spectroscopic aperture typically covers most of a galaxy’s starlight, resulting in an absorption profile that is an ensemble measurement of thousands of individual outflowing clouds. 
A complementary technique to counter the loss of positional information is to probe the disk-CGM interface via QSO absorption spectroscopy. 

Combining down-the-barrel observations with QSO-absorption spectroscopy for the same galaxies will allow us to trace the outflow from the energy injection zone (few pc) out to the CGM (tens of kpc). 
A space-based narrow slitlet high-resolution multi-object spectrograph will allow us to use local star-forming galaxies and target spatially resolved star clusters individually, enabling the characterization of outflows associated with each region. Combining this with QSO or background galaxy spectroscopy at the disk-CGM interface will allow us to trace the outflowing material as it moves away from the galaxy.

Some of the lines of interest are the coronal lines such as \OVI, \NV, and \CIV\ with ionization potential between 40–140~eV that are uniquely suited to trace the wind-CGM interactions, while lower and intermediate ionization species (e.g., \SiII, \SiIII, \CII, \MgII, \FeII, etc.) are suited to trace the bulk of the outflowing material. The far and near-UV wavelength ranges give access to these lines that can trace the complex multiphase nature of the outflows.  

These observations will require high spatial resolution (sub-arcsec) to localize the injection point, continuous coverage from far-UV to near-UV coverage from 1000 to 3500~\AA, and high spectral resolution of (R~$\ge$~50,000) to identify individual clouds. 
These design requirements will facilitate the accurate measurement of mass outflow rates, as well as the properties of the outflow, including metallicity, kinematics, ionization state, and energetics, as well as the microphysics of the wind-CGM interactions.

\subsubsection{Fermi Bubbles and AGN-driven feedback}

AGN-driven nuclear winds can deposit substantial amounts of energy in the CGM of galaxies \citep{Faucher-Giguere12}.
Understanding and characterizing these nuclear winds is an important part of the broader effort to understand the Galactic ecosystem and baryon cycle.

In the MW, clear evidence for the nuclear wind is provided by the Fermi Bubbles {\color{black}(or eROSITA bubbles)}, giant plasma lobes extending more than 10 kpc on both sides of the Galactic center. 
The bubbles emit radiation across the spectrum from gamma rays \citep{Su2010, Ackermann2014}, X-rays \citep{Predehl2020}, mid-IR emission \citep{BH03} to microwaves and radio waves \citep{Finkbeiner2004, Carretti2013}.
Further evidence for the nuclear wind is provided by high-velocity clouds (HVCs) seen in 21 cm emission \citep{MG13, dT18, LockmanJ2020} and UV absorption  \citep{Keeney2006, Fox2015, Bordoloi2017, Savage2017, Karim2018, Ashley2020, Ashley2022} in sightlines passing through the Fermi Bubbles. 
These Fermi Bubble HVCs trace cool and warm ($T\!\sim\!10^{4-5}$ K) gas clouds embedded in the nuclear wind.
Fermi Bubble counterparts are frequently seen in galaxy simulations \citep{Pillepich2021}, but observational evidence for counterparts in other galaxies is sparse. 
A gamma-ray structure of size 6--7.5 kpc is tentatively seen in M31, on either side of its center \citep{Pshirkov2016}. Searches for Fermi Bubble counterparts are needed across a larger sample of galaxies.

With its far-UV capabilities and improved sensitivity, HWO will be able to observe many more QSOs and halo stars behind the Fermi Bubbles than the Hubble Space Telescope (HST), particularly in the poorly understood low-latitude regions of the Bubbles.
HWO will also probe similar structures in other nearby galaxies, by observing galaxy/QSO pairs at small impact parameters of $<$20\,kpc. With a larger statistical sample, HWO will be able to characterize the prevalence of such nuclear bubbles, their sizes, and physical properties to quantify the role of AGN-feedback in the nearby Universe.

\newpage
\renewcommand{\arraystretch}{1.}
\begin{table}[]
\begin{longtable}{|p{2.0cm}|p{3.0cm}|p{11.0cm}|}
\caption{Lines of interest and their associated phenomenon}\\
\hline
\textbf{Transitions} & \textbf{Wavelength (\AA)} & \textbf{Importance }\\  
\hline
H{\sc i}$^{a}$    & Lyman series (1215 -- 920) & Traces neutral gas content at low and intermediate temperatures ($T = 10^{1-5}~K$). The most common set of absorption lines.\\
Lyman~$\alpha^{a,e}$    &          1215   &       Strongest emission line tracing neutral hydrogen and the bulk of the cool gas mass in the CGM.\\
C{\sc ii}$^{a}$  & 1036,1334 &One of the strong metal-lines suitable for probing low metallicity or low column density gas at low-ionization states.  \\
C{\sc iii}$^{a,e}$   & 977& One of the strong metal-lines suitable for probing low metallicity or low column density gas at intermediate ionization states. {\color{black}Expected to be one of the strongest} metal emission lines.\\
C{\sc iv}$^{a}$  & 1548, 1550 &  Strong lines tracing higher ionization states. Together with O{\sc vi}  and N{\sc v} can be a powerful tool to identify non-equilibrium processes. Also, known to trace interactions of galactic winds and the CGM. \\
N{\sc v}$^{a}$  &  1238, 1242  & Coronal lines with powerful diagnostic power to identify young (1-5 Myr) stellar population and non-equilibrium processes. \\
O{\sc i}$^{a}$  &  1025, 1039, 1302 & Strong tracer of primarily neutral gas and is coupled to neutral hydrogen. \\
O{\sc vi}$^{a,e}$  &  1031, 1038   & Tracer {\color{black} of (1) gas at the viral temperature of 0.1--1~L$_{\star}$ galaxies, (2) energetic interactions expected between the cool CGM clouds and hot winds/outflows, (3) photoionized gas; and (4) non-equilibrium processes. Suitable for high resolution (R$>$50,000) absorption and low resolution (R$\approx$5,000) emission spectroscopy. One of the most powerful high-ionization diagnostic transitions.} \\
Si{\sc ii}$^{a}$  &  1190,1193,1260, 1304,1526 & Commonly observed species with five lines of different intrinsic strength that enable characterization of the saturation level and hence accurate estimation of column densities.\\
Si{\sc iii}$^{a}$  &   1206 & Most common metal-line seen in absorption in the halos of galaxies. Together with Si{\sc ii} and Si{\sc iv} allows for {\color{black}accurately measuring} of the ionization state of the gas.\\
Si{\sc iv}$^{a}$  &   1393,1402 & Together with Si{\sc ii} and Si{\sc iii} enables {\color{black}accurately measuring} the ionization state of the gas.\\
S{\sc ii}$^{a}$  & 1250,1253,1259 & Weak lines associated with a low-depletion species that enable accurate metal measurement for high column density systems like damped \Lya\ systems.\\
Mg{\sc ii}$^{a,e}$  & 2796, 2803 & Commonly traced species associated with extended gas disks and cool outflowing material. Extensive literature on this line at z$>$0.5. Expected to be detectable in emission. \\
H$_2 ^{a}$ & bands from 1150--912& Traces the bulk of the cold molecular gas. Critical for accurate estimation of the baryon budget and gas flows that may support short-term star formation.\\
\hline
\end{longtable}
\vspace{0.2cm}
\noindent $^a$ lines suitable for absorption-line spectroscopy\\
$^e$ lines suitable for emission-line spectroscopy\\
\end{table}

\section{Physical Parameters and Measurements \label{sec:physical_parameters}} 

To make significant improvements in all the avenues discussed above and to catapult our understanding of how galactic ecosystems are regulated, we advocate that HWO have a multi-object spectrograph capable of making absorption and emission measurements at high spatial and spectral resolution with continuous coverage from far-UV to near-UV.

HWO should enable astronomers to --
\begin{itemize}
    \item[1.] Measure the above-mentioned properties for a statistical sample of multiple QSO and background galaxy sightlines per target galaxy to counter CGM stochasticities (\S\ref{sec:HWO_obstype}). This would require HWO to be sensitive to measure the continuum of background sources with apparent magnitude of 21 mag (AB magnitude) in the GALEX FUV band at sufficient signal-to-noise $\approx 10$ or higher (see \S\ref{HWO-sensitivity}).
    \item[2.] Measure robust column densities for gas in low, intermediate, and high ionization states via absorption spectroscopy using background targets. These would require measuring cold and cool species such as H$\rm _2$, \HI, \SiII, \CII, \MgII, \FeII, \SII\ and others, intermediate-ionization species such as \SiIII, \CIII, and \SiIV, and high-ionization species such as \OVI, \NV, \CIV, and \NeVIII. Measuring these species would require a minimum wavelength coverage from 970--3000 \AA\ (see \S\ref{HWO-wavelength}).
    \item[3.] Measure the kinematics and component structure of individual species to distinguish individual clouds and the dominant physical process(es) responsible for the state of the gas. We recommend a resolution of R$\sim 100,000$, allowing a physical velocity separation of 3~\kms\ (see \S\ref{HWO-resolution}). 
    \item[4.] Measure emission around the bright UV background sightlines probing the absorbing clouds. Having both emission and absorption measurements will provide a direct estimate of the density of the medium \citep{Dixon2001,Shelton2001}. The emission measurements may cover a larger spatial region and at a lower spectral resolution of R$\gtrsim$5000 tracing strong emission lines such as \Lya, \OVI, \CIII, \Lyb, \CIV, \MgII, and others (see \S~\ref{HWO-MOSsetup}).
\end{itemize}

\renewcommand{\arraystretch}{1.5}
\begin{longtable}{|p{3.0cm}|p{3.0cm}|p{3.0cm}|p{3.0cm}|p{3.0cm}|}
\caption{State-of-the-art and requirements for progress}\\
\hline
\textbf{Physical Parameter} & \textbf{State of the Art} & \textbf{Incremental Progress (Enhancing)} & \textbf{Substantial Progress (Enabling)} & \textbf{Major Progress (Breakthrough)} \\
\hline
\endfirsthead

\hline
\textbf{Physical Parameter} & \textbf{State of the Art} & \textbf{Incremental Progress (Enhancing)} & \textbf{Substantial Progress (Enabling)} & \textbf{Major Progress (Breakthrough)} \\
\hline
\endhead

1. Column density and covering fraction of gas clouds around galaxy disks (beyond Milky Way) & Sensitivity to absorbers with column densities of HI $\sim10^{13.5-14}$ cm$^{-2}$ separated by about 20~\kms.
About 20 galaxies probed by sightlines within 30 kpc. & Sensitivity to absorbers with column densities of HI $\sim10^{13-13.5}$ cm$^{-2}$ separated by 20~\kms. About five galaxies probed by at least two sightlines within 30~kpc. & Sensitivity to absorbers with column densities of HI $\sim10^{13-13.5}$ cm$^{-2}$ separated by 10~\kms. About 25 galaxies probed by three or more sightlines within 30 kpc. & Sensitivity to absorbers with column densities of HI $\sim10^{13}$ cm$^{-2}$ separated by 5~km~s$^{-1}$. About 25 galaxies probed by four to six sightlines within 30~kpc. This is motivated by predicted covering fractions, e.g., the covering fraction of HVCs is $\approx 100\%$ within 30 kpc. \\
\hline
2. The kinematics of gas clouds in and around the disks at the disk-CGM interface of galaxies in the low-z Universe ($z\sim0.01$) & Kinematics of the order of 20~\kms\ tracing large-scale outflows or inflows in active systems. Data on MW and $\sim100$ other galaxies. & Kinematics at a resolution of 10~\kms\ tracing fountain flows and gas inflows with multiple sightlines around galaxies. & Kinematics at 5--7~\kms\ to identify individual components in outflowing and inflowing gas & Kinematics at 3--5~\kms\ for point and extended sources. {\color{black} This will provide the ability to isolate individual clouds associated with narrow absorption components in the outer disk tracing low dispersion medium ($\rm \sigma \sim 6~kms^{-1}$ \citep{Sellwood99}) seen in \HI\ 21cm studies.}\\
\hline
3. Ionization state and chemical composition (metallicity) of the gas & Highly ionized gas traced by OVI and cooler components traced by higher order Lyman series lines, H$_{2}$ in the UV with limited archival FUSE data for low-z systems. These data established the presence of hot and cold gas in the disk-CGM interface of the MW and other nearby galaxies. & High ionization phase of gas traced by OVI, intermediate ionization gas traced by CIV, SiIV, and lower ionization state traced by HI, SiII, SiIII, etc. This will enable constraining the ionization state and multi-phase nature of the gas. & Access to Lyman series lines, molecular hydrogen lines as well as OVI, CIV, SiIV, SiII, SiIII, etc., spanning from $\sim970-1650$ \AA. & Access to Lyman series lines, molecular hydrogen lines as well as OVI, CIV, SiIV, SiII, SiIII, MgII, ZnII, and CrII spanning from $\sim970-3000$ \AA, enabling measurement of ionization state, metallicity, and dust content. \\
\hline
4. Sizes of clouds and covering fractions & Large-scale structures of $\sim100$~kpc with QSO sightlines with HST and ground-based observatories. {\color{black}Sightlines through the MW probe sub-kpc spatial scales.} & Probing $<50$~kpc scale clouds that most photoionization models predict. Having about 2--3 sightlines within 20 kpc of the galaxy disk with coverage of FUV and NUV species. & Probing $<$10~kpc scale clouds containing the bulk of neutral gas in CGM. 6 sightlines within 20 kpc. & Sub-kpc scales probing structure in clouds. 2-3 extended background sources (galaxies) per target galaxy allowing multiple sightlines within 20 kpc. \\
\hline
5. Density of the clouds/medium & One set of measurements obtained for the MW when combining emission measurements from FUSE with absorption measurements (spatially apart). Extragalactic systems lack such measurements. & Extending emission and absorption studies to the same region of the sky for a half a dozen targets. & Extending emission and absorption studies to the same region of the sky for a couple of dozen targets. Measurements for 4-6 extragalactic targets. & Emission and absorption studies using hundreds of QSO sightlines tracing MW CGM and covering 30-40\% of the sky. Extending emission and absorption studies to tens of extragalactic targets. \\
\hline
\end{longtable}

\section{HWO Performance Requirements \label{sec:HWO_performance}}

\subsection{Type of Observations \label{sec:HWO_obstype}}
 The primary observational technique we highlight here is absorption line spectroscopy, which traces small-scale gas properties as absorption against a bright background source. QSOs, galaxies, and stars with large  UV fluxes may be targets for observations. Absorption studies can be either of the following:
\begin{itemize}
    \item[1.] Pencil-beam studies using background point sources like QSOs or stars. Using multiple sightlines, one can gather statistics on covering fractions and the distribution of the gas in the disk-CGM interface.
    \item[2.] Absorption studies with extended UV-bright background sources. This strategy has the advantage of mapping the small-scale structure of the clouds at sub-kpc sampling in projection over several-kpc scales.

\end{itemize}
We also advocate measuring CGM in emission via intrinsically strong lines in the vicinity of the absorbing system. The emission may be measured at lower spatial (tens of kpc scales) and spectral resolution (R $\sim$ 5000) than the absorption measurements. These measurements, when combined with absorption measurements, will enable estimation of the densities and sizes of clouds. These parameters are essential in inferring the origin, lifetimes, and fate of the clouds and the processes involved in their creation and destruction.

While measuring individual sightlines will provide detailed information on individual systems, stacking or binning, on the other hand, can be employed to reveal the statistical properties of a sample for which individual high S/N measurements are beyond the capability of HWO.
Therefore, the design of the spectrograph must be such that binning or stacking the data should reduce noise and enhance S/N. Detector technologies like photon-counting devices that do not add significant noise will be best suited for achieving this critical requirement.

\subsection {Potential Targets and HWO Sensitivity \label{HWO-sensitivity}}

We highlight three sets of potential targets that will enable HWO to produce a detailed map of gaseous structures in the disk-CGM interface in absorption.

{ \color{black} The first would be a scaled-up version of current studies with sample sizes that are 5 to 10 times larger. A sample of more than 200 low-z galaxies probed by distant background UV-bright targets will be possible with HWO. Such a large sample will enable characterizing the disk-CGM interface as a function of galaxy properties and stochasticities within galaxy populations. }

\begin{figure}
\centering
\includegraphics[width=1\linewidth]{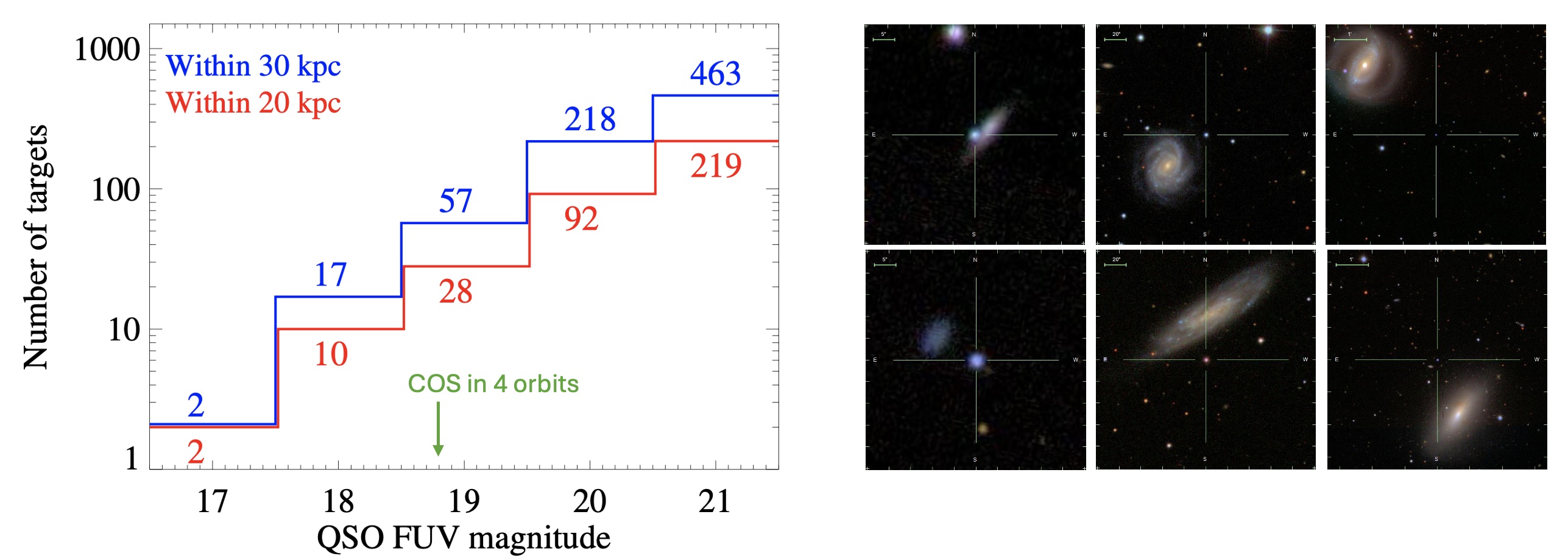}
\caption {Left: Number of galaxy-pairs as a function of UV magnitude of the background QSO. The redshift range of the foreground galaxies was chosen to be between z=0.002 (recession velocity of 600~\kms) and z=0.1 (systems where \OVI\ and \Lyb\ are not easily accessible {\color{black} by} HST).  The red and the blue lines indicate QSOs within 20 and 30 kpc of the foreground galaxy, respectively. The pairs were obtained by cross-matching GALEX Ultraviolet Sources \citep{galex_source_catalog} with the spectroscopic galaxy catalog from the Sloan Digital Sky Survey {\color{black}\citep{Abolfathi:2018aa}} While the numbers might change if using other galaxy catalogs due to differences in their sky coverages, spectroscopic completeness, and other biases, the general trend is expected to be the same.  {\color{black} The sensitivity of the state-of-the-art spectrograph, COS, is shown as a green arrow. The COS can achieve a S/N=10 for a limiting FUV magnitude of 18.8 in 4 orbits at 1250 \AA\ where its sensitivity is optimal. At this sensitivity, only a dozen sources are {\color{black} available} to COS. However, the wavelength coverage of Cosmic Origins Spectrograph (COS) makes O{\sc vi}, C{\sc iii}, and higher order Lyman series lines for these low-z galaxies inaccessible. } Right: Examples of galaxy-QSO pairs with QSOs ranging from 19 -- 21 mag in the FUV within 20~kpc of the target galaxies. The QSO sightline is positioned at the center of the image, and the galaxy is on the side. The scale is indicated at the top {\color{black} left } corner and shows 5$^{\prime\prime}$, 20$^{\prime\prime}$, and 1$^{\prime}$ from the leftmost to rightmost column.}
\label{fig:Sample_FUV_QSO}
\end{figure}

Figure~\ref{fig:Sample_FUV_QSO} shows the possible samples of galaxy QSO pairs from z=0.002--0.1. As expected, the sample sizes can be significantly improved if HWO provides sensitivity to 21 mag or fainter QSOs. We note that while many more pairs are available if we consider z$<$0.001 galaxies, absorption spectroscopy of such systems suffers from confusion with MW lines. There are multiple ways to counter the confusion if high sensitivity and resolution are available.

{\color{black} The second would be to observe a dozen unique galaxies where we have multiple QSO sightlines (3--5 sightlines) within 20~kpc of individual galaxies. Signatures of various physical processes within a single galaxy will manifest as variations in the CGM properties among the sightlines. 
For example, the Fermi Bubble occupies 10\% of the area within 20~kpc \citep{Su2010, Dobler10} of the Milky Way, and star formation-driven outflows also occupy 20--35\% of the area depending on the opening angle \citep{Schneider20}. Therefore, we expect outflows and Fermi Bubble-like structures to have a spatial covering fraction of 20--35\% of the disk-CGM interface. This requires sampling individual galaxies with 3-5 sightlines to capture the signatures of specific feedback processes.}
To {\color{black}account for} stochasticity, we need a sample of 6--10 galaxies per luminosity bin. Based on the QSO luminosity function, the sensitivity of HWO should enable observations of 22 FUV AB magnitude targets {\color{black}to provide large enough samples and do so} at S/N $\sim$10 in reasonable exposure times.

\begin{figure}
\centering
\includegraphics[width=1\linewidth]{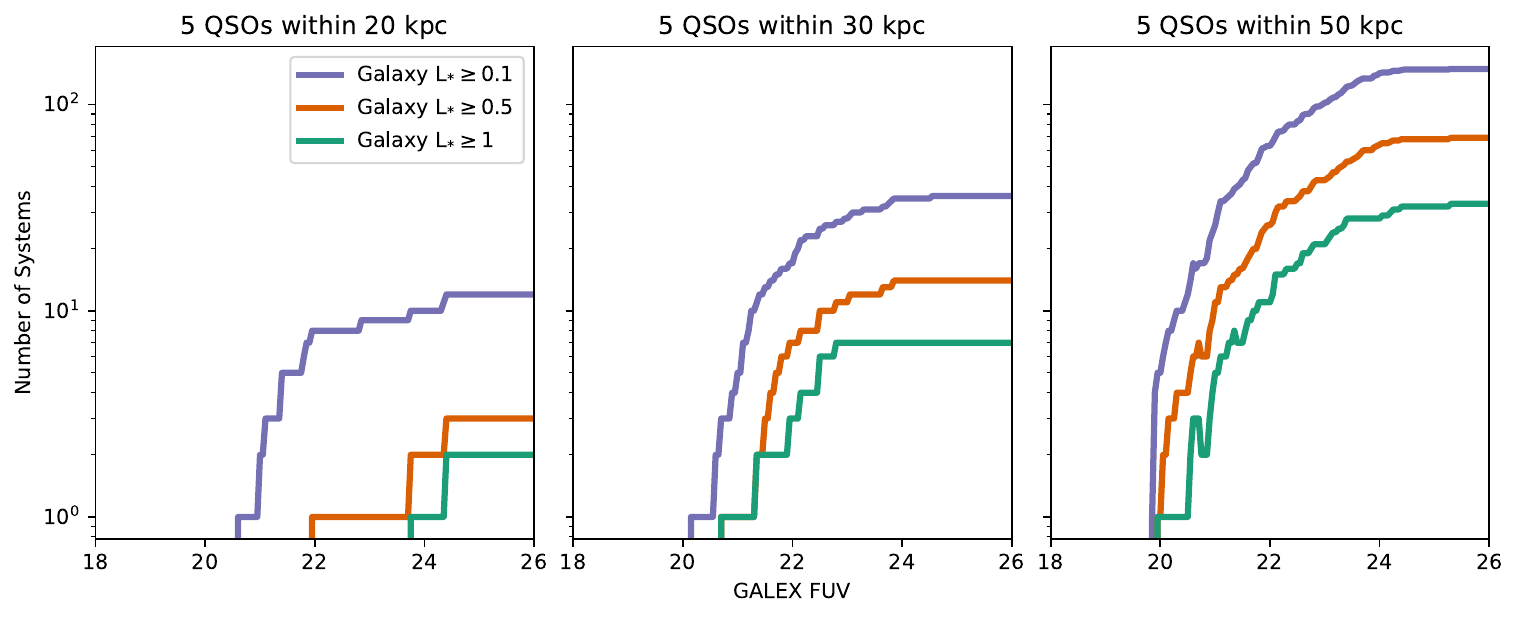}
\caption{The number of galaxies with at least five background QSOs within 20, 30, and 50 kpc is shown as a function of limiting QSO GALEX FUV magnitude. Minimum foreground galaxy luminosities of $\rm 0.1~L_{\star}, ~0.5~L_{\star}, and ~1~L_{\star}$ are shown by the purple, orange, and green lines, respectively. Background QSOs are drawn from the Milliquas survey \citep{Flesch2023Million} cross-matched with the Revised Catalog of GALEX Ultraviolet Sources \citep{galex_source_catalog}, and foreground galaxies are drawn from the NASA Extragalactic Database.}
\label{fig:50kpc_source_density}
\end{figure}

Figure~\ref{fig:50kpc_source_density} shows predictions for the number of possible targets with multiple sightlines per galaxy within 20, 30, and 50~kpc for a range of limiting magnitudes of the background QSOs. Adding background galaxies can significantly improve the number of targets, although the Lyman continuum break in the background galaxy's spectrum will limit the redshift range for selecting them for absorption spectroscopy.

{\color{black} The third set of potential targets would be far-UV bright sources with a relatively ``absorption feature-free" spectrum that probes the disk-CGM interface of the MW. There will be hundreds of sources satisfying this criterion, which can be further screened to target specific MW features on the sky (e.g., Fermi Bubble, Magellanic bridge, HVCs, etc.).
All extragalactic sightlines will enable absorption measurements of the MW disk-CGM interface along with the extended MW halo and the local group medium at kpc or sub-kpc scales. Disk and halo stars will probe more local regions in the disk-CGM interface. Better spectra (with higher S/N and resolution) for targets previously observed by FUSE and HST will enhance our models of the MW disk-CGM interface. }

In conclusion, {\color{black} the sample sizes can be dramatically improved if HWO can reach a limiting FUV magnitude of 21 or higher with a S/N=10 down to 1000\AA\ in a 4-hr exposure. 
This will result in a significantly higher density of potential background sources accessible to HWO than currently for the HST-COS.
COS is limited to 18.8~mag QSOs to acquire a medium-resolution spectrum with the G130M grating for a 4-orbit (4$\times$50~min) exposure. COS would take 50~orbits (155~ksec) to observe a 21~mag source at an S/N=10, R=20000, and 1250 \AA\ where COS is most sensitive. Therefore, such observations for a statistical survey are prohibitively expensive with HST-COS and highlight the need for the HWO.}

\subsection{Wavelength Coverage \label{HWO-wavelength}}

We recommend that HWO cover the bluer end of the far-UV wavelength as close to the Lyman limit as possible. Access to higher-order Lyman series transitions would substantially benefit all science cases described here. With access to \Lyb\ $\lambda 1025$ and \Lyg\ $\lambda 972$, substantial progress can be made in estimating neutral hydrogen column densities, which is critical for gas budget, metallicity measurement, ionization corrections, and estimating physical properties of the clouds. Access to \Lyd\ by going down to 940\AA\ would be transformative as this will enable measurements of \HI\ column densities for individual clouds using the optically thin transitions. 

\begin{figure}
\label{fig:wavelength_coverage}
    \centering
    \includegraphics[width=\linewidth]{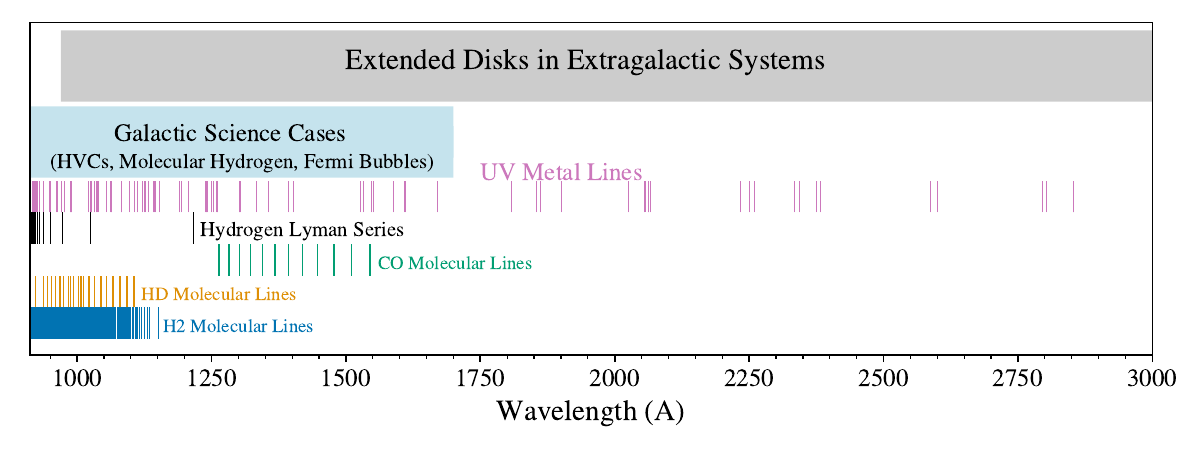}
    \caption{Distribution of common UV lines critical for studying the disk-CGM interface. Gray and blue shades highlight the UV wavelength coverages needed to achieve the corresponding science cases as discussed in Section 2. Magenta lines show UV metal lines from ions commonly detected in {\color{black}the} CGM/IGM, such as \SiII, \SiIII, \SiIV, \CIV, \OVI, and \MgII. Black lines show hydrogen Lyman series, with the leftmost approaching the Lyman limit at 912~\AA. Green, brown, and blue lines show CO, HD, and $H_2$ molecular lines, respectively. }
\end{figure}

Figure~\ref{fig:wavelength_coverage} illustrates the rest-frame UV coverage needed to target specific lines and highlights the science topics that will specifically require that coverage. It highlights the need {\color{black}for covering} the far and extreme UV wavelengths $<$1000~\AA\ to access multiple lines from metal species, molecular hydrogen, and high-order hydrogen lines.

The blue end will allow one to measure three critical metal lines -- \OVI $\lambda$1031, \CIII$\lambda$977, and \OI $\lambda$1039-- tracing highly ionized, intermediate, and primarily neutral gas. 
The wavelength range from 912--1150\AA\ is also critical to detect many ro-vibrational lines of H$_2$, including the Werner and Lyman bands, tracing the bulk of the molecular gas in the disk-CGM interface.

The high sensitivity coverage of the rest of the far-UV (1150--1700 \AA) is also critical to access multiple transitions including \SII $\lambda\lambda 1250, 1253, 1259$, \SiII $\lambda\lambda 1190, 1193, 1260, 1304$, \SiIII $\lambda1206$, \NV$\lambda 1238, 1242$, \SiIV$\lambda\lambda 1393, 1402$, \CIV\ $\lambda$1548, 1550, which are essential lines for metallicity measurements and for modeling ionization states.  

The rest-frame near-UV ($\sim$2000-3000 \AA) coverage is also key to determining the metallicity and dust content of the CGM disk interface using powerful tracers like the Zn II  $\lambda\lambda \rm 2026, 2062$~\AA, Cr~II $\lambda\lambda \rm 2056, 2062, 2066$~\AA, and Mg~II $\lambda \lambda \rm 2796,2803$~\AA. 
A large body of literature exists on higher redshift systems in these transitions; however, very little is known about the low redshift universe (z$<$ 0.5 ) covering the last 5 billion years.

 In summary, we advocate for continuous coverage from  {\color{black} 940 to 3500~\AA\ } to enable potentially transformative science by opening a new spectral window for the nearby Universe ($\rm z=0 - 0.25$). A minimum wavelength range of {\color{black} 970 to 3000 \AA} is necessary to realize the scientific advances recommended by the Astro2020 decadal survey.

\subsection{Resolution \label{HWO-resolution}}

Spectral resolution is a key consideration for studies of the disk-CGM interface. The processes active in this region produce structures with a kinematically dense signature. In {\color{black} current observations}, component structures in the absorbing material can be seen down to the resolution limit of the spectrograph, indicating that we could be missing rich kinematic information. In addition, blending of components would also severely interfere with our ability to measure gas column densities and other properties.

A resolution of 100,000 (3~\kms) would be ideal and will guarantee that lines are detected beyond the confusion limit due to blending.
The component structure of gas in absorption in the disk-CGM interface regions becomes crowded. Data from the HST and ground-based observatories have shown that the component structure can only be moderately resolved at R=30,000 {\color{black} but} well-resolved at $R$=100,000. 

{\color{black} Cool ``ISM-like" neutral clouds at temperatures of 100--1,000~K \citep{Wolfire03} } have thermal widths of a few \kms (2--6~km/s for Hydrogen or 4 and 5 times smaller for Oxygen and Silicon, respectively). 
For example, the expected full-width half max {\color{black} for the three S~II lines ($\lambda \lambda 1250, 1253, 1259$)}, critical for estimating metallicity, is $1.2$~\kms\ for a temperature of 1,000~K. {\color{black} Therefore, line blending due to thermal width is expected only in the case of extremely high resolution data at R$>$ 300,000. }

Realistically, gas can be broadened both thermally and non-thermally. Turbulence can be an important factor in line-broadening; however, precise measurements are needed to ascertain that. A resolution of $\ge$~100,000 will ensure that individual kinematic components from the MW disk and halo can be resolved and measured for multiple ions. We present two examples that support the need for resolution greater than 50,000 and preferably 100,000.

\begin{figure}
    \centering
    \includegraphics[scale =.54]{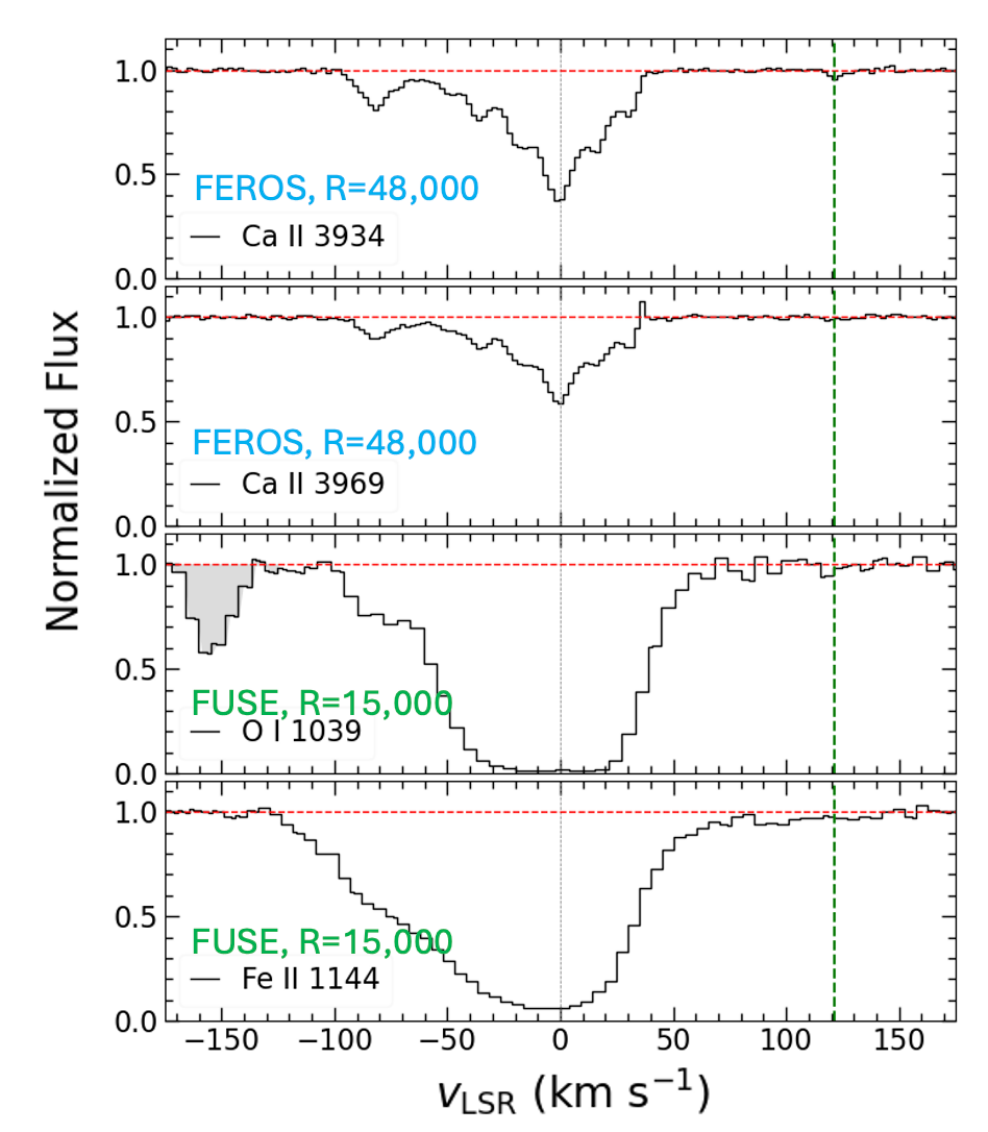}  
    \caption{Velocity profiles of MW ISM absorption in optical and ultraviolet wavelengths seen in the foreground of Galactic Center star HD 156359. In both panels, the normalized flux is shown in black, and the continuum level is in red. The vertical line at 0 \kms\ marks the region associated with the Milky Way. Left: Ca{\sc ii} $\lambda\lambda$3934, 3969 optical lines in an archival $R=48,000$ spectrum taken with ESO/FEROS \citep{cashman2023}. Right: O{\sc i} $\lambda$1039 and Fe{\sc ii} $\lambda$1144 ultraviolet lines in an archival $R=15,000$ spectrum taken with the FUSE spectrograph \citep{cashman2023}. Regions shaded in gray are unrelated to the transition. The green vertical line near $+120$ \kms\ marks the location of an HVC in Complex WE, which is visible in the optical FEROS spectrum (for the stronger $\lambda$3934 line), but not in the FUV FUSE spectrum.}
     \label{fig:res_hires_vs_fuse}
\end{figure}

Figure~\ref{fig:res_hires_vs_fuse} compares the same absorbing clouds in two sets of observations -- optical spectra from the ground using ESO/FEROS and UV spectra with FUSE (data reproduced from \citep{Cashman2021}). 
At the resolution of FUSE (R=15,000), the high-velocity components of \OI\ and \FeII\ are blended, whereas at a resolution of R=48,000, the optical spectra show complex component structure in \CaII. At low resolution, the absorption lines tend to get smeared and lost in the noise, as in the case of the feature marked by the green dotted line. Blending of components has also been inferred for COS data probing the disk-CGM interface, as indicated by a strong correlation between the Doppler b parameter and column density \citep{Koplitz25}. This trend is enhanced in species with a single transition, where situation levels are not directly estimated.

\begin{figure} 
    \centering
    \includegraphics[width=\linewidth]{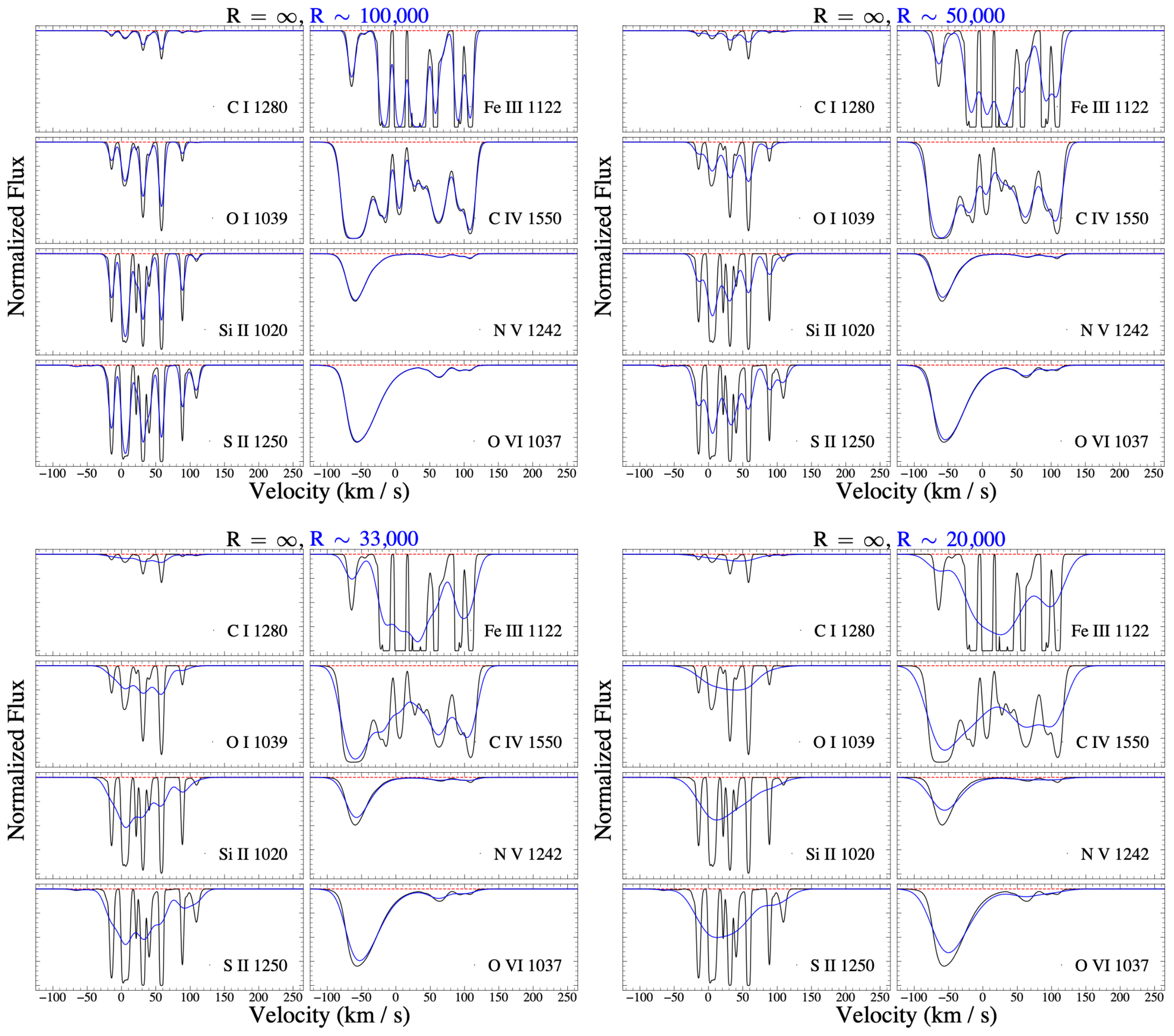}
    \caption{Synthetic sightlines created from the MAIHEM simulations \citep{Koplitz23}. These simulate observing a halo of mass $10^{12}\ M_\odot$ at an impact parameter of $\sim$16 kpc. At R=100000, the spectral features are recovered completely, although some of the narrow saturated lines may look unsaturated. The dilution of spectral features progresses with the drop in spectral resolution, which is noticeable at R=50,000 where information on levels of saturation of lines is lost. By R=33000, significant degradation can occur. Some narrow, densely packed features get completely lost and component structure is fully blended i.e., irrecoverable. Interestingly, this is the resolution for the state-of-the-art studies with HST/COS. Improving resolution to at least 50,000 is necessary for significant progress in the field.}
    \label{fig:resolution}
\end{figure}

We have also investigated the role of resolution in blending the features using {\color{black}a} theoretical study \citep{Koplitz23} of the CGM driven by turbulence. {\color{black} Figure~\ref{fig:resolution} shows the infinite resolution synthetic spectra from the simulated disk-CGM interface of a galaxy of halo mass $10^{12}\ M_\odot$ at an impact parameter of 16~kpc in black and the same at various other resolutions in blue.} A comparison between the two indicates that R=100,000 can perfectly resolve the component structure, whereas R=50,000 is the minimum resolution required to guarantee the detection of the individual lines. At the resolution of our current most sensitive UV spectrograph, the Cosmic Origins Spectrograph (COS) with R=20,000, most of the component structure at the disk-CGM interface will be irretrievably lost.

\subsection{Aperture Requirements \label{HWO-aperture}}

{\color{black}The stringent requirements for high resolution and high S/N within a reasonable exposure time require that the HWO must have an aperture significantly larger than that of the HST. The precise aperture size would depend on the design of the telescope and properties of the components (coatings, detector efficiencies, etc.), which will ultimately determine the throughput and effective area of the instrument. In the absence of those details, we present a back-of-the-envelope calculation here for the minimum aperture size.

We assume that advances in mirror coatings, detectors, and telescope design will lead to five times better efficiency in photon capture for HWO than for HST. This assumption implies that for the same aperture size and observing time, HWO will be able to acquire the same data quality (S/N) at $R$=100,000 as HST at $R$=20,000. Since we require HWO to achieve an S/N of 10 for a 21 mag target in 4~hrs (14.4~ksec) instead of 155~ksec, the aperture size must be at least $\sqrt{155/14.4}= 3.3$ times that of HST or 8\,m in diameter. If throughput drops by 25\% at the bluer wavelengths, then to achieve the same efficiency, HWO should have an aperture of 9\,m in diameter. 
These zeroth-order calculations match the models presented in the HWO exposure time calculator (\url{https://hwo.stsci.edu/uvspec_etc}), although it does not include wavelengths shorter than 1000~\AA.}

\subsection{Instrumental mode \label{HWO-MOSsetup}}

{\color{black} The two main requirements that dictate the design of the instrument are the field of view (FOV) and the spatial size of the slitlet. For a galaxy in the local Universe (10~Mpc away), an FOV of $\rm 6^{\prime} \times 6^{\prime}$ would cover a region of the galaxy $\rm \approx 20~kpc \times 20~kpc$. This matches the height above the disk where we expect the processes discussed above to be most active. On the other hand, a slitlet size of $<1^{\prime\prime}$ or $\sim 50~pc$ will be needed to precisely identify gas flows associated with star clusters/associations. In principle, these requirements can be met by an integral field spectrograph (IFS) or a multi-object spectrograph (MOS). However, a full-coverage IFS would need to have more than 200,000 spaxels. Coupled with the requirement for high spectral resolution and large wavelength coverage, an IFS may not be feasible. However, a MOS with thousands of slitlets that can be placed/opened at the desired positions within the FOV would be efficient.}

{\color{black} A MOS will be an efficient design, as the detector space can be utilized for the sky positions of interest. It will enable simultaneous observations of multiple background sources tracing the CGM as well as individual star-forming regions within the stellar body of the galaxy to trace gas flows.}
Sub-arcsec spatial resolution of the MOS will also enable astronomers to use extended targets as background sources to map the gas in absorption at parsec scales. A combination of observing galaxy-QSO pairs and galaxy-galaxy pairs will provide a detailed map of the clouds in the disk-CGM interface, tracing extended disks, HVCs, outflows, and AGN-driven bubbles like the Fermi Bubble.

\begin{figure}
    \centering
    \includegraphics[width=\linewidth]{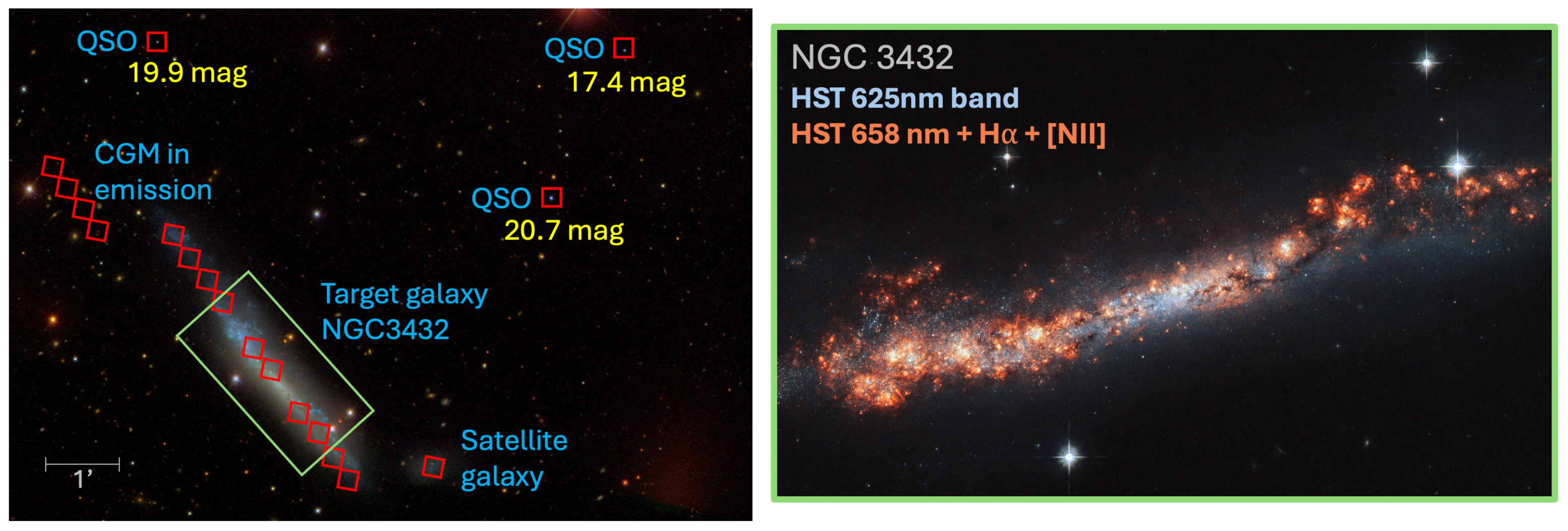}
    \caption{(Left) We show a MOS mode setup around NGC 3432 that covers the QSO sightlines, the stellar body of the galaxy, nearby satellite galaxies, and the CGM of NGC 3432. The red squares denote 1$^{\prime\prime}$ slitlets of the MOS (not to scale) and the yellow circles mark the position of the background UV-bright QSOs. A similar setup can be imagined for other nearby galaxies. At a distance of 12~Mpc, a MOS with $6^{\prime} \times 6^{\prime}$ FOV will cover a region of 22~kpc $\times$ 22~kpc for NGC~3432. 
    (Right:) HST r-band (blue) and H$\alpha +$[NII] (orange) composite image of NGC~3432 showing the star forming regions along the disk. While such galaxies have been well studied with deep imaging with HST and other facilities, our understanding of their CGM is sparse. Being one of the galaxies with three QSOs brighter than 21 mag within 25~kpc from its disk, NGC3432 provides a unique opportunity to probe the variations in the properties of the disk-CGM interface while acquiring information on the stellar populations within the disk and nearby satellite galaxies.   }
    \label{fig:instrument_setup}
\end{figure}

A preferred observation design would be a MOS that enables simultaneous observations of thousands of spatial positions. MOS or microshutter array (MSA) designs with small spatial slitlets ($< 1^{\prime\prime}$) and a large field of view of a few arcminutes square (e.g., $6^{\prime} \times 6^{\prime}$) will enable simultaneous measurement of spectra from the target galaxy as well as background sources. Such a design will allow the disk-CGM interface to be connected to the properties of the host galaxy. An example of a possible MOS setup (not to scale) is presented in Figure~\ref{fig:instrument_setup} for the star-forming galaxy NGC~3432 with three UV-bright background QSOs. While some of the slitlets can be placed on the background QSOs and brighter parts of the galaxy to detect the foreground gas in absorption, other ``star free" slitlets can be placed in stellar continuum-free regions to detect the CGM in emission. Data from multiple stellar continuum free slitlets when stacked (averaging multiple positions) and binned (reducing spectral resolution) may enable the detection of the disk-halo interface in emission.

We {\color{black} now} summarize the requirements for HWO compared to the state-of-the-art and the level of improvement.

\renewcommand{\arraystretch}{1.0}

\begin{longtable} {|p{3cm}|p{3cm}|p{3cm}|p{3cm}|p{3cm}|}
\caption{Observational capabilities necessary for progress in the field}\\
\hline
\textbf{Observation Requirement} & \textbf{State of the Art} & \textbf{Incremental Progress (Enhancing)} & \textbf{Substantial Progress (Enabling)} & \textbf{Major Progress (Breakthrough)} \\
\hline
\endfirsthead
\hline
\textbf{Observation Requirement} & \textbf{State of the Art} & \textbf{Incremental Progress (Enhancing)} & \textbf{Substantial Progress (Enabling)} & \textbf{Major Progress (Breakthrough)} \\
\hline
\endhead

Spectroscopy and low-resolution imaging &  {\color{black} Spectroscopy at R$\sim$15,000 for background targets available with HST-COS. Higher resolution R$\sim$100,000 with HST-STIS only for the brightest targets. Both are single aperture/slit designs. Imaging is not available in the far-UV.} & {\color{black} High-resolution spectroscopy R$\sim$30,000 (10~km/s) for multiple (tens) targets in the FOV with higher throughput to reach 19--20 mag targets.} & {\color{black} High-resolution spectroscopy R$\sim$50,000 (6~km/s). MOS design with hundreds of slitlets for simultaneous observation for mapping spatial variations.} & {\color{black} Very high-resolution spectroscopy R$\sim$100,000 (3~km/s) with thousands of slitlets. The slitlet positions can be used to mimic imaging.} \\
\hline
Wavelength Range & FUV wavelength coverage for spectroscopy from {\color{black}1150}--1700~\AA. & FUV wavelength coverage for spectroscopy down to 1000~\AA~and extending NUV coverage to 2000~\AA. & Contiguous coverage from FUV to optical from 1000--2800 \AA. & Contiguous coverage from extreme-UV to optical, i.e.,  from 940--3500~\AA. \\
\hline
Magnitude of target in chosen bandpass & Spectroscopic capabilities to detect sources with FUV mag 19 at S/N $\sim$10 in 4~hours. & Spectroscopic capabilities to detect sources with FUV mag 20 at S/N$\sim$10 in 4~hours. & Spectroscopic capabilities to detect sources with FUV mag 20.5 at S/N$\ge$10 in 4~hours. & Spectroscopic capabilities to detect sources with FUV mag 21 at S/N$\ge$10 in 4~hours . \\
\hline
FOV & Single pointing $3^{\prime\prime}$ on the sky. & Multiobject Spectrograph with each slitlet of size $<3^{\prime\prime}$ and a FOV $\approx 1^{\prime} \times 1 ^{\prime}$ & Multiobject Spectrograph with  100s of slitlets of $1^{\prime\prime}$ and a FOV $\approx 3^{\prime} \times 3^{\prime}$. & Multiobject Spectrograph with at least a 1000 slitlets of subarcsec resolution with an FOV of $\gtrsim 6^{\prime} \times 6^{\prime}$. \\
\hline
\end{longtable}

\section{Conclusion \label{sec:conclusion}}

The disk-CGM interface is a unique region that holds the signature of gas flows and the energetics of feedback processes. However, because of its lower densities, it has yet to be explored thoroughly. In this science case, we advocate for access to absorption spectroscopy in a contiguous spectral range from extreme-UV through far-UV to near-UV that covers a rich diversity of lines tracing a variety of ionic species. Absorption-line spectroscopy will not only enable detection of the faint gas otherwise not accessible for imaging, but is also independent of the distance to the target resulting in a statistically unbiased sampling.

We recommend that the next generation UV telescope, the HWO, should have a sensitive high-resolution multi-object spectrograph with the following characteristics:

\begin{itemize}
    \item [1.] Sensitivity to detect 21 mag FUV target with S/N $\le$10 in a few hours;
    \item [2.] Covers a wavelength range from 940--3500 \AA\ (minimum of 970--3000 \AA);
    \item [3.] Provides a spectral resolution of R$\approx$ 100,000 but no less than 50,000;
    \item [4.] Provides a FOV $\ge 6^{\prime} \times 6^{\prime}$ with at least a thousand sub-arcsec slitlets; and 
    \item [5.]  Has noise properties such that binning and stacking the high-resolution data results in enhanced S/N.
\end{itemize}

{\color{black} This design will also enable the detection of the gas in the disk-CGM interface in emission when added over multiple slitlets and spectral pixels. The reduction in spatial and spectral resolution would enhance S/N, which would make emission line detections possible. Although an IFS design may be more suitable for emission studies, the MOS would mimic a patchy IFS that would be tunable in terms of spectral and spatial resolution.}

The HWO, a flagship NASA mission, is poised to revolutionize our understanding of the pathways of gas flow that maintain galactic ecosystems. The observations and instrument design requirements discussed here will enable astronomers to answer some of the critical questions related to baryon cycling in galaxies and provide a new window to galactic ecosystems in the nearby Universe.

\newpage

% \disclosures 
\subsection*{Disclosures}
The authors have no relevant financial interests in the manuscript or other potential conflicts of interest to disclose.

\subsection* {Code, Data, and Materials Availability} 
Data and plots presented in the article are obtained from publicly available sources, and the citations are provided in the text.

\subsection* {Acknowledgments}
The authors thank the referees for their constructive comments that have significantly improved this work.
The authors also acknowledge the input of the community through the HWO IGM-CGM working group. 
SB and BK acknowledge the support from grant HST-GO-14071 administered by STScI and NASA ADAP grant No. 80NSSC21K0643.

%%%%% References %%%%%

\bibliography{report}   % bibliography data in report.bib
\bibliographystyle{spiejour}   % makes bibtex use spiejour.bst

%%%%% Biographies of authors %%%%%

\vspace{2ex}\noindent\textbf{Sanchayeeta Borthakur} is an Associate Professor at Arizona State University. She received her Ph.D. degree in astronomy from the University of Massachusetts, Amherst, in 2010. She has authored/co-authored more than 50 journal articles. Her research interests include UV spectroscopy, studies of the CGM, the IGM, and the ISM. She has led multiple programs with the HST and the VLA, and is the co-chair of the IGM/CGM science working group for the HWO.  

\vspace{2ex}\noindent\textbf{Joseph N. Burchett} is an Assistant Professor at New Mexico State University. He received his Ph.D. in astronomy from the University of Massachusetts, Amherst, in 2017.  Burchett is also co-chair of the IGM/CGM science working group for HWO and has led several programs on HST. His scientific interests include galaxy evolution as it relates to the environment and the role played by the IGM and CGM.

\vspace{1ex}\noindent\textbf{Frances Cashman} is an Assistant Professor at Presbyterian College. She received her Ph.D. in Physics from the University of South Carolina in 2020. She is interested in UV spectroscopy, ISM/CGM studies of the Milky Way--Local Group, and laboratory astrophysics. 

\vspace{1ex}\noindent\textbf{Andrew Fox} is an ESA-AURA Astronomer at STScI in Baltimore, where he leads the Milky Way Halo research group. He received his Ph.D. in Astronomy from the University of Wisconsin--Madison in 2005. He studies diffuse gas in the ISM and CGM, particularly in the Milky Way and Local Group. He has led international collaborations with the HST, VLT, and GBT.

\vspace{1ex}\noindent\textbf{Yong Zheng} is an Assistant Professor at Rensselaer Polytechnic Institute. She received her Ph.D. degree in Astronomy from Columbia University in 2018. She is interested in the cosmic baryon cycle that connects galaxies and their CGM, from both observation and simulation perspectives. Her research extensively uses UV spectroscopy and HI 21cm observations (e.g., GALFA-HI, HI4PI).

\vspace{1ex}\noindent\textbf{David M. French} is a Staff Scientist at STScI in Baltimore, MD. He received his Ph.D. in Astronomy from the University of Wisconsin--Madison in 2018. He is interested in the IGM and CGM, with an emphasis on how the gaseous environments of galaxies influences their growth and evolution.

\vspace{1ex}\noindent\textbf{Rongmon Bordoloi} is an Associate Professor of Physics at North Carolina State University. His primary research focus is understanding the evolution of galaxies by studying the extended gaseous halos around them. He and his team utilize cutting-edge ground-based instruments such as Keck, VLT, and Magellan, alongside space-based telescopes like HST and JWST.

\vspace{1ex}\noindent\textbf{Brad Koplitz} is an Astrophysics Ph.D. candidate at Arizona State University. He is interested in using spectroscopy to study the CGM and IGM, particularly the ionization processes taking place.

\vspace{1ex}\noindent Biographies and photographs of the other authors are not available.

\listoffigures
\listoftables

\end{spacing}
\end{document}